\documentclass[letterpaper,twocolumn,fleqn]{article}

\usepackage{ist}
\usepackage{amsmath,amsfonts,amssymb}
\usepackage{booktabs}
\usepackage{multirow}
\usepackage{colortbl}
\usepackage[colorlinks=true,linkcolor=blue,citecolor=blue,urlcolor=blue]{hyperref}
\usepackage{xcolor}
\usepackage{subcaption}
\usepackage{enumitem}
\usepackage{placeins}
\usepackage{stfloats}

\title{Multimodal Fusion of Skeleton Dynamics and Clinical Gait Features for Video-Based Cerebral Palsy Severity Assessment}

\author{Kaiyuan Yang$^{a}$, Xupeng Chen$^{b}$, Jiangpeng He$^{a}$\\[4pt]
$^{a}$Department of Computer Science, Luddy School, Indiana University Bloomington, Bloomington, USA\\
$^{b}$Department of Electrical and Computer Engineering, Tandon School, New York University, Brooklyn, USA\\
Corresponding author: jhe2@iu.edu}

\date{}

\begin{document}

\maketitle

\thispagestyle{empty}

%%%%%%%%%%%%%%%%%%%%%%%%%%%%%%%%%%
% Abstract
%%%%%%%%%%%%%%%%%%%%%%%%%%%%%%%%%%

\begin{abstract}
Video-based gait analysis has become a promising approach for assessing motor impairment in children with cerebral palsy (CP). However, existing methods usually rely on either pose sequences or handcrafted gait features alone, making it difficult to simultaneously capture spatiotemporal motion patterns and clinically meaningful biomechanical information. To address this gap, we propose a multimodal fusion framework that integrates skeleton dynamics with contribution-guided clinically meaningful gait features. First, Grad-CAM analysis on a pre-trained ST-GCN backbone identified the most discriminative body keypoints, providing an interpretable basis for subsequent gait feature extraction. We then build a dual-stream architecture, with one stream modeling skeleton dynamics using ST-GCN and the other encoding gait geatures derived from the identified keypoints. By fusing the two streams through feature cross-attention improved four-level CP motor severity classification to 70.86\%, outperforming the baseline by 5.6 percentage points. Overall, this work suggests that integrating skeleton dynamics with clinically meaningful gait descriptors can improve both prediction performance and biomechanical interpretability for video-based CP severity assessment.

\end{abstract}

% \smallskip
\noindent\textbf{Keywords:} Cerebral Palsy, Gait Analysis, Spatiotemporal Graph Convolutional Networks, Multimodal Fusion, Clinical Interpretability

%%%%%%%%%%%%%%%%%%%%%%%%%%%%%%%%%%
% Introduction
%%%%%%%%%%%%%%%%%%%%%%%%%%%%%%%%%%
\vspace{10pt}
\section{Introduction}
\label{sec:intro}

Cerebral palsy (CP) is one of the most common childhood motor disability, affecting approximately 2--3 per 1,000 live births worldwide~\cite{rosenbaum2007report}, with substantial regional variation in prevalence~\cite{olusanya2022cerebral} and significant economic burden on healthcare systems~\cite{batista2025costs}. Because gait dysfunction is a core manifestation of CP, scalable and objective video-based gait assessment has become an important direction for screening, follow-up, and treatment evaluation.

Traditional clinical assessment remains widely used~\cite{vitrikas2020cerebral}, but it is often subjective and depends on expert judgment, which may introduce inter-rater variability and limit sensitivity to subtle changes over time. Instrumented gait analysis can provide detailed biomechanical measurements~\cite{haberfehlner2020instrumented}, but it usually requires specialized laboratories and expensive motion capture sensor systems, which limits its accessibility. Caregiver-reported questionnaires can reflect daily functional impact~\cite{bautista2018psychometric}, but they are also indirect and may not provide sufficiently objective measurements.

Recent advances in computer vision have made video-based gait assessment a promising alternative~\cite{kidzinski2020deep}, addressing accessibility and cost barriers in motor functional disease management. A typical pipeline extracts 2D keypoints from walking videos via pose estimation (e.g., OpenPose~\cite{cao2017realtime}), models spatiotemporal dynamics, and maps them to clinically meaningful outcomes. Kidzi\'{n}ski et al.~\cite{kidzinski2020deep} demonstrated that clinically relevant gait features can be inferred from monocular videos. Zhao et al.~\cite{zhao2024computer} applied ST-GCN with metric learning for CP severity prediction, while Deng et al.~\cite{deng2024interpretable} demonstrated interpretable video-based tracking for motor-state quantification in Parkinson's disease, highlighting the broader applicability of markerless video assessment.

Despite these advances, existing video-based methods still face an important gap. Skeleton-based models are effective at capturing spatiotemporal motion patterns directly from pose sequences, but they mainly use raw keypoint trajectories so that they lack of interpretability~\cite{halilaj2018machine}, and do not explicitly incorporate the biomechanical descriptors that are commonly used in clinical gait analysis. In contrast, clinical gait features (e.g., joint angles, step length, symmetry indices) are interpretable and clinically grounded, Samadi~Kohnehshahri et al.~\cite{samadi2024machine} reported that many machine learning approaches for CP gait data provide limited transparency, and Xiang et al.~\cite{xiang2025explainable} highlighted joint-angle and symmetry-related descriptors as key interpretable signals. These findings motivate methods that connect models with clinically meaningful variables. However the low-dimensional summaries of clinical gait features may miss rich temporal information. As a result, these two sources of information are often modeled separately, even though they are naturally complementary. 

\begin{figure*}[t]
    \centering
    \includegraphics[width=\textwidth]{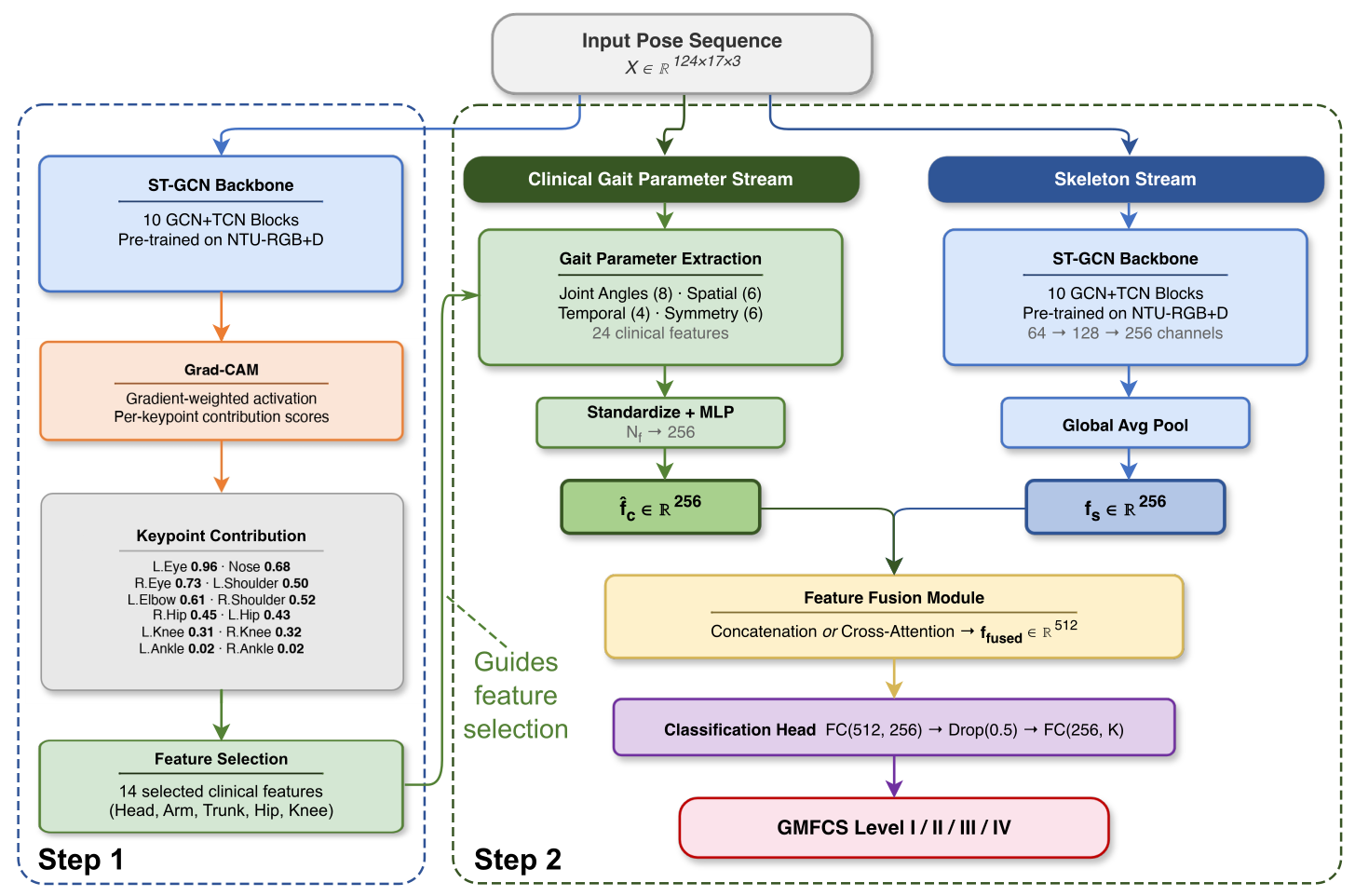}
    \caption{Overview of the proposed two-step framework. \textbf{Step~1}: An ST-GCN backbone is pre-trained and analyzed via Grad-CAM to identify discriminative keypoints, guiding gait feature selection. \textbf{Step~2}: A dual-stream architecture combines skeleton dynamics (ST-GCN backbone $\to$ global average pooling $\to$ $\mathbf{f}_s \in \mathbb{R}^{256}$) with extracted clinical gait features (MLP projection $\to$ $\hat{\mathbf{f}}_c \in \mathbb{R}^{256}$) via feature fusion for GMFCS classification.}
    \label{fig:architecture}
\end{figure*}

In this work, to address the mentioned gap between two different modalities, we propose a two-step multimodal fusion framework for video-based CP severity assessment (Fig.~\ref{fig:architecture}). In \textbf{Step~1}, we apply Grad-CAM~\cite{selvaraju2017grad} to a pre-trained ST-GCN backbone to identify the body keypoints that contribute most strongly to prediction. These keypoints are then used to guide the extraction of clinically informed gait features. In \textbf{Step~2}, we build a dual-stream architecture consisting of a skeleton stream for modeling spatiotemporal dynamics and a clinical feature stream for encoding gait features derived from the selected keypoints. The two streams are fused through feature cross-attention. In this study, CP severity is formulated as a four-class GMFCS label~\cite{palisano1997development} for supervised classification, which is a common standard in CP motor function assessment to evaluate the performance of models. The main contributions of this work are threefold. First, we introduce a Grad-CAM-guided feature selection strategy that links neural network interpretability with clinically informed gait feature engineering. Second, we propose a dual-stream fusion framework that integrates skeleton dynamics with clinical gait features, leading to a 5.6 percentage point improvement in classification accuracy over the baseline model. Third, we provide a systematic comparison of feature selection and fusion strategies, showing that clinically guided feature subsets can be more effective than using a large set of features without selection.

%%%%%%%%%%%%%%%%%%%%%%%%%%%%%%%%%%
% Related Work
%%%%%%%%%%%%%%%%%%%%%%%%%%%%%%%%%%
\section{Related Work}
\label{sec:related}

\textbf{Video-Based Gait Assessment for CP.}
Recent advances in markerless pose estimation have made video-based gait assessment increasingly feasible for children with CP. OpenPose~\cite{cao2017realtime} and HRNet~\cite{sun2019deep} have been widely used for this purpose, providing the pose sequences needed for downstream gait analysis. Prior work has shown that clinically relevant gait parameters can be inferred from monocular video. Kidzi\'{n}ski et al.~\cite{kidzinski2020deep} pioneered single-camera prediction of clinical gait metrics, and Zhao et al.~\cite{zhao2024computer} further used ST-GCN with metric learning for CP severity prediction (with GMFCS as an evaluation label). Le and Pham~\cite{le2024learning} proposed a Transformer-based network for estimating gait parameters from single-view RGB videos, while Arias~Valdivia et al.~\cite{ariasvaldivia2025deep} applied deep learning to postural control analysis for CP subtype classification.

\textbf{Graph Convolutional Networks for Skeleton Analysis.}
ST-GCN~\cite{yan2018spatial} introduced spatial graph convolutions combined with temporal convolutions for skeleton-based action recognition, and later variants improved representation power through adaptive topologies~\cite{shi2019two}, channel-wise topology refinement~\cite{chen2021channel}, and attention mechanisms~\cite{plizzari2021spatial}. These methods are effective for modeling motion dynamics, but most focus on latent skeleton representations without explicitly encoding biomechanical descriptors. Jun et al.~\cite{jun2023hybrid} showed that combining skeleton features with gait parameters benefits pathological gait recognition, and Jing et al.~\cite{jing2023deep} reported clinically meaningful accuracy for deep learning-assisted gait parameter assessment. Together, these works support a multimodal direction that jointly leverages temporal dynamics and structured clinical features.

\textbf{Interpretability in Clinical Gait Analysis.}
Clinical adoption of deep learning for gait analysis remains limited by insufficient interpretability~\cite{halilaj2018machine}. Samadi~Kohnehshahri et al.~\cite{samadi2024machine} reported that many machine learning approaches for CP gait data provide limited transparency, and Xiang et al.~\cite{xiang2025explainable} highlighted joint-angle and symmetry-related descriptors as key interpretable signals. These findings motivate methods that connect model attention with clinically meaningful variables. Our framework addresses this need by combining Grad-CAM-based keypoint attribution (Step~1) with explicit clinical gait features (Step~2).

%%%%%%%%%%%%%%%%%%%%%%%%%%%%%%%%%%
% Methods
%%%%%%%%%%%%%%%%%%%%%%%%%%%%%%%%%%
\section{Methods}
\label{sec:methods}

\subsection{Overview of the Two-Step Framework}

Our framework operates in two steps (Fig.~\ref{fig:architecture}):

\textbf{Step~1 (Keypoint Importance Analysis):} A pre-trained ST-GCN backbone is analyzed using Grad-CAM~\cite{selvaraju2017grad} to compute per-keypoint contribution scores, revealing which body joints the network relies upon for GMFCS classification. These scores guide the selection of clinically meaningful gait features from the most discriminative body regions.

\textbf{Step~2 (Dual-Stream Fusion):} The input pose sequence is processed by two parallel streams---a skeleton stream and a clinical gait parameter stream---whose outputs are integrated via a feature fusion module for GMFCS classification.

The input to both steps is a 2D pose sequence $\mathbf{X} \in \mathbb{R}^{T \times V \times C}$ with $T{=}124$ frames, $V{=}17$ COCO keypoints, and $C{=}3$ channels (x, y, confidence).

\subsection{Step 1: ST-GCN Backbone and Keypoint Importance}
\label{sec:step1}

\subsubsection{ST-GCN Architecture}

We adopt the ST-GCN architecture~\cite{yan2018spatial,zhao2024computer} as our skeleton backbone. The COCO 17-keypoint skeleton is represented as a spatial graph $\mathcal{G} = (\mathcal{V}, \mathcal{E})$.

Each ST-GCN block consists of: Spatial Conv2d $\to$ Spatial Graph Multiplication (with the adjacency matrix) $\to$ BatchNorm2d $\to$ ReLU $\to$ Temporal Conv2d $\to$ BatchNorm2d $\to$ Dropout $\to$ Residual connection $\to$ ReLU. The backbone passes the input through Data BatchNorm1d, then 10 cascaded ST-GCN blocks (with channel dimensions progressing $64 \to 128 \to 256$), followed by global average pooling to produce the skeleton embedding $\mathbf{f}_s \in \mathbb{R}^{256}$.

The backbone is initialized with weights pre-trained on NTU-RGB+D~\cite{shahroudy2016ntu} for action recognition, providing rich spatiotemporal motion representations.

\subsubsection{Grad-CAM Keypoint Contribution Analysis}

To understand which body joints the ST-GCN backbone considers most important for GMFCS classification, we apply Grad-CAM~\cite{selvaraju2017grad} to the final convolutional layer. Grad-CAM computes gradient-weighted activation maps that highlight the spatial regions (keypoints) contributing most to the classification decision.

\begin{figure}[t]
    \centering
    \includegraphics[width=0.85\columnwidth]{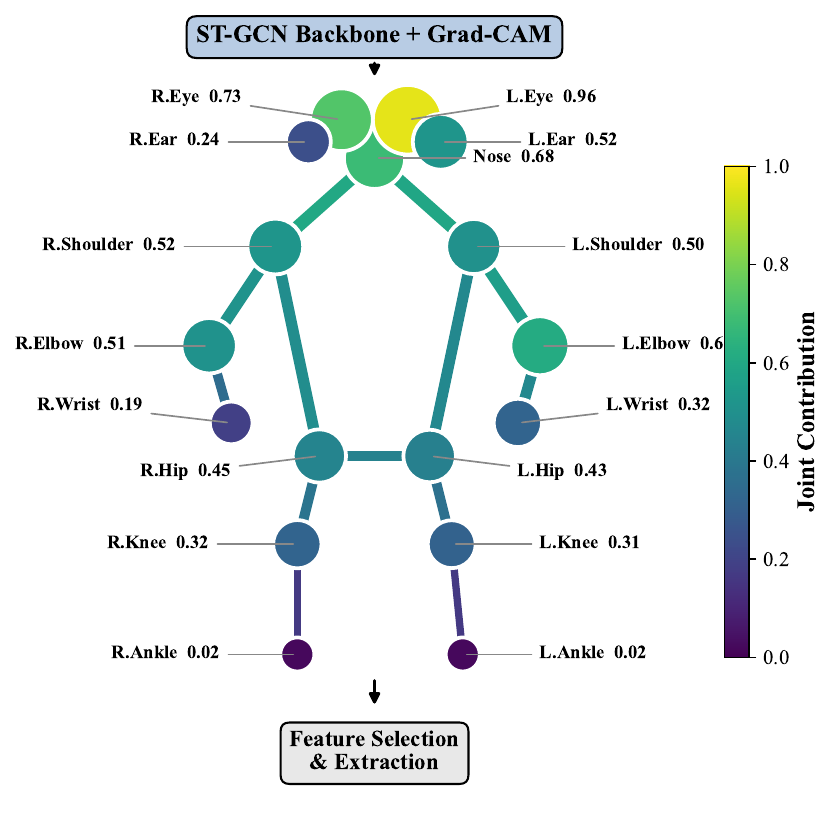}
    \caption{Grad-CAM keypoint contribution analysis on the ST-GCN backbone. Joint circle sizes and colors indicate contribution scores to GMFCS classification. Head and upper-body keypoints show the highest contributions, while ankles contribute minimally. These scores guide the selection of body regions for clinical gait feature extraction.}
    \label{fig:gradcam}
\end{figure}

Fig.~\ref{fig:gradcam} shows the resulting per-keypoint contribution scores. Several observations guide our feature selection:
\begin{itemize}[leftmargin=*,nosep]
    \item \textbf{Head region} (eyes, nose) shows the highest contributions (0.68--0.96), reflecting the importance of head posture and control in CP severity assessment.
    \item \textbf{Upper body} (shoulders, elbows) shows moderate-to-high contributions (0.50--0.61), capturing arm swing and trunk stability.
    \item \textbf{Lower body} (hips, knees) shows moderate contributions (0.31--0.45), encoding gait kinematics.
    \item \textbf{Ankles} show the lowest contributions (0.02), likely because 2D ankle positions are less informative than the angular and temporal features computed from them.
\end{itemize}

These findings motivate our clinical gait feature extraction: we extract features from five body regions---\textbf{Head}, \textbf{Arm}, \textbf{Trunk}, \textbf{Knee}, and \textbf{Hip}---that correspond to keypoints with moderate-to-high Grad-CAM contribution scores.

\subsection{Step 2: Dual-Stream Fusion Architecture}
\label{sec:step2}

\subsubsection{Skeleton Stream}

The skeleton stream uses the same ST-GCN backbone described in Section~\ref{sec:step1}. The output spatiotemporal feature map is globally pooled to produce $\mathbf{f}_s \in \mathbb{R}^{256}$.

\subsubsection{Clinical Gait Feature Extraction}
\label{sec:gait_params}

Guided by the Grad-CAM analysis, the clinical stream extracts biomechanical features from five body regions: \textbf{Head} (neck angle from ear--shoulder--hip vectors), \textbf{Arm} (arm-body angle, swing amplitude), \textbf{Trunk} (inclination, lateral body sway), \textbf{Hip} and \textbf{Knee} (joint angles via the law of cosines, range of motion, bilateral symmetry). Frame-wise joint angles are computed as:
\begin{equation}
    \theta_{\mathrm{joint}}(t) = \arccos\!\left(\frac{\mathbf{a}(t) \cdot \mathbf{b}(t)}{|\mathbf{a}(t)|\,|\mathbf{b}(t)|}\right)
\end{equation}
We additionally extract temporal features (cadence, gait cycle duration, stance-swing ratio) and spatial features (step length, stride length, walking speed). Bilateral symmetry indices~\cite{robinson1987use} are computed as $\mathrm{SI} = |X_L - X_R| / [0.5(X_L + X_R)] \times 100\%$.

In total, 24 features are extracted. We compare all 24 features against a clinically curated subset of 14 features (Table~\ref{tab:features}), selected based on Grad-CAM guidance and clinical relevance, excluding features with high redundancy or low 2D reliability. All features are standardized and processed by a two-layer MLP to produce $\hat{\mathbf{f}}_c \in \mathbb{R}^{256}$.

\subsubsection{Feature Fusion Strategies}
\label{sec:fusion}

We explore two strategies for integrating skeleton and clinical feature embeddings:

\textbf{Concatenation Fusion.}
The simplest approach directly concatenates the two embeddings:
\begin{equation}
    \mathbf{f}_{\mathrm{fused}} = \mathbf{f}_s \oplus \hat{\mathbf{f}}_c \in \mathbb{R}^{512}
\end{equation}

\textbf{Cross-Attention Fusion.}
A sigmoid-gated attention mechanism computes a dynamic weighting:
\begin{equation}
    \alpha = \sigma\!\left(\frac{\mathbf{f}_s^\top \hat{\mathbf{f}}_c}{\sqrt{d}}\right), \quad d = 256
\end{equation}
\begin{equation}
    \mathbf{f}_{\mathrm{fused}} = \mathrm{LN}\!\left(\mathbf{f}_s + \alpha \hat{\mathbf{f}}_c\right) \oplus \mathrm{LN}\!\left(\hat{\mathbf{f}}_c + \alpha \mathbf{f}_s\right)
\end{equation}
where $\mathbf{f}_{\mathrm{fused}} \in \mathbb{R}^{512}$.

The fused representation is passed to a classification head: FC$(512 \to 256)$ $\to$ Dropout $\to$ FC$(256 \to K)$ for GMFCS prediction.

\begin{table}[!t]
\centering
\caption{Selected clinical gait features (14 features) organized by body region. Features were selected based on Grad-CAM keypoint importance and clinical relevance.}
\label{tab:features}
\vspace{3pt}
\begin{tabular}{@{}llr@{}}
\toprule
\textbf{Body Region} & \textbf{Features} & \textbf{Count} \\
\midrule
Hip     & L/R hip ROM, hip ROM sym.  & 3 \\
Knee    & L/R knee ROM, knee ROM sym. & 3 \\
Trunk   & Trunk inclination & 1 \\
Spatial & Step len., stride len., speed & 3 \\
Temporal & Cadence, gait cycle dur. & 2 \\
Symmetry & Step len. sym., timing sym. & 2 \\
\bottomrule
\end{tabular}
\end{table}

\subsection{Training Strategy}
\label{sec:training}

\textbf{Phase~1 (epochs 1--3):} ST-GCN backbone frozen; only the fusion module and classifier are trained with a learning rate of $10^{-3}$.
\textbf{Phase~2 (epochs 4--20):} Last two ST-GCN blocks unfrozen for joint fine-tuning with a reduced learning rate of $10^{-4}$.
We use cross-entropy loss, Adam optimizer (weight decay $5{\times}10^{-5}$), cosine annealing scheduler, batch size 128, and data augmentation (horizontal flip with left/right keypoint swap, Gaussian noise $\sigma{=}2$\,px). Model selection is based on the best validation accuracy, and early stopping is applied with a patience of 5 epochs. All experiments are implemented in PyTorch and conducted on a single NVIDIA GPU.

%%%%%%%%%%%%%%%%%%%%%%%%%%%%%%%%%%
% Experiments
%%%%%%%%%%%%%%%%%%%%%%%%%%%%%%%%%%
\section{Experiments}
\label{sec:experiments}

\subsection{Dataset}

We evaluate on the public CP gait dataset from Kidzi\'{n}ski et al.~\cite{kidzinski2020deep}, collected at Gillette Children's Specialty Healthcare (1994--2015). The dataset contains gait videos from 1,026 patients with CP across GMFCS Levels~I--IV, recorded using consumer-grade cameras during clinical visits. OpenPose~\cite{cao2017realtime} extracts 25 keypoints per frame, which are converted to COCO 17-keypoint format. Sliding window sampling (window$=$124 frames, stride$=$12) with quality filtering ($\geq$80\% keypoints with confidence$\geq$0.2) yields 2{,}946 clips from GMFCS Levels~I--IV. We use a patient-level stratified split (Training: 1{,}574, Validation: 473, Test: 899) to prevent data leakage. Table~\ref{tab:dataset} summarizes the dataset distribution.

\begin{table}[t]
\centering
\caption{Dataset distribution across GMFCS levels and data splits.}
\label{tab:dataset}
\begin{tabular}{@{}lrrrr|r@{}}
\toprule
 & \textbf{I} & \textbf{II} & \textbf{III} & \textbf{IV} & \textbf{Total} \\
\midrule
Train & 556 & 597 & 390 & 31 & 1{,}574 \\
Val   & 167 & 179 & 118 & 9 & 473 \\
Test  & 314 & 327 & 244 & 14 & 899 \\
\midrule
Frac. & 35.3\% & 37.9\% & 24.8\% & 1.6\% & 2{,}946 \\
\bottomrule
\end{tabular}
\end{table}

GMFCS Level~IV is severely under-represented (1.6\%), reflecting the lower prevalence and fewer clinical visits at this severity level in the source cohort.

\subsection{Main Results}

Table~\ref{tab:results} presents the classification performance across all evaluated configurations on the 899-sample test set. We report accuracy, weighted F1 score (accounting for class imbalance), and linearly weighted Cohen's kappa ($\kappa_l$), which is appropriate for ordinal classification.

\begin{table}[t]
\centering
\caption{GMFCS classification performance on the test set ($N{=}899$). Best in \textbf{bold}. ``Sel.'' denotes the clinically curated 14-feature subset.}
\label{tab:results}
\begin{tabular}{@{}p{2.6cm}rrr@{}}
\toprule
\textbf{Method} & \textbf{Acc} & \textbf{F1$_w$} & \textbf{$\kappa_l$} \\
\midrule
ST-GCN Baseline & 65.29 & 0.658 & 0.597 \\
+ All Gait (XAttn)     & 67.41 & 0.670 & 0.619 \\
+ Sel.\ Gait (XAttn)   & 69.08 & 0.682 & 0.646 \\
\textbf{+ Sel.\ Gait (Concat)} & \textbf{70.86} & \textbf{0.706} & \textbf{0.665} \\
\bottomrule
\end{tabular}
\end{table}

\textbf{Key observations.}
(1)~All fusion variants outperform the skeleton-only baseline, confirming the value of clinical gait parameters.
(2)~Feature selection matters: the Grad-CAM-guided 14-feature subset outperforms all 24 features by 1.7\% under cross-attention, indicating that domain-informed curation reduces noise from redundant features on moderately sized clinical datasets.
(3)~The best configuration---selected features with concatenation---achieves \textbf{70.86\% accuracy} (+5.6~pp over baseline), with gains in weighted F1 (+0.048) and $\kappa_l$ (+0.068).
For context, Zhao et al.~\cite{zhao2024computer} reported 76.6\% on a private dataset (not directly comparable); Kidzi\'{n}ski et al.~\cite{kidzinski2020deep} focus on gait parameter prediction rather than GMFCS classification.

\subsection{Per-Class Analysis}

Table~\ref{tab:per_class} shows per-class recall (sensitivity) for the baseline and best configuration.

\begin{table}[t]
\centering
\caption{Per-class recall (\%) by GMFCS level. $N$ denotes the number of test samples.}
\label{tab:per_class}
\begin{tabular}{@{}lrrrr@{}}
\toprule
\textbf{Method} & \textbf{I} & \textbf{II} & \textbf{III} & \textbf{IV} \\
 & \scriptsize{$N{=}314$} & \scriptsize{$N{=}327$} & \scriptsize{$N{=}244$} & \scriptsize{$N{=}14$} \\
\midrule
ST-GCN  & 63.69 & 61.77 & 74.18 & 28.57 \\
\textbf{Ours} & \textbf{69.43} & \textbf{63.61} & \textbf{86.07} & 7.14$^\dagger$ \\
\midrule
$\Delta$ & +5.74 & +1.84 & +11.89 & --- \\
\bottomrule
\end{tabular}

\vspace{2pt}
{\scriptsize $^\dagger$Level~IV contains only 14 test samples; per-class metrics at this level are not statistically reliable.}
\end{table}

The largest improvement occurs at GMFCS Level~III (+11.9\%), where clinical gait abnormalities---reduced cadence, pronounced asymmetry, increased joint angle deviations---are most salient and well-captured by the clinical feature stream. Level~I also shows notable gains (+5.7\%).

\subsection{Confusion and ROC Analysis}

\begin{figure*}[!t]
\centering
\includegraphics[width=\textwidth]{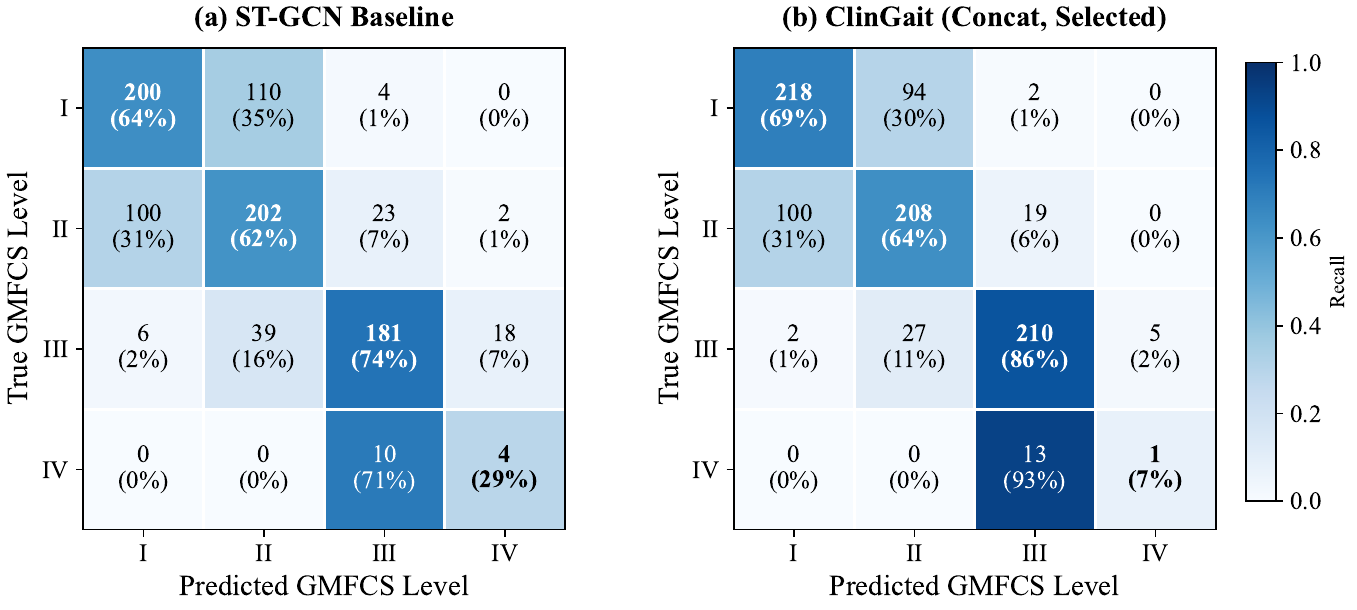}
\caption{Confusion matrices for (a) ST-GCN baseline and (b) our framework (selected features, concatenation). Cell values show raw counts and row-normalized recall percentages.}
\label{fig:confusion}
\end{figure*}

\begin{figure*}[!t]
\centering
\includegraphics[width=\textwidth]{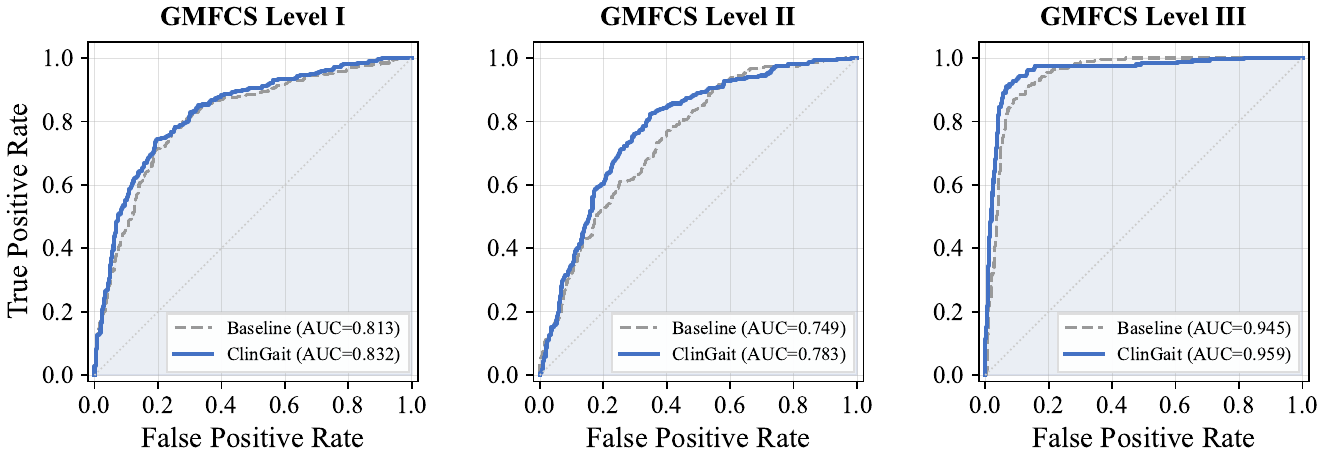}
\caption{Per-class ROC curves for GMFCS Levels~I--III, comparing ST-GCN baseline (dashed gray) and our framework (solid blue). AUC values are shown in the legend.}
\label{fig:roc}
\end{figure*}

Fig.~\ref{fig:confusion} shows the confusion matrices. The baseline exhibits substantial adjacent-level confusion, particularly between Levels~I and~II (210 misclassifications). Our framework reduces this to 194 and improves Level~III recall from 74.2\% to 86.1\% by reducing III$\to$II errors (39$\to$27). Fig.~\ref{fig:roc} presents per-class ROC curves for Levels~I--III (Level~IV omitted due to $N{=}14$). Our framework improves AUC across all levels, with the largest gain on Level~II (0.749$\to$0.783). Level~III achieves the highest AUC (0.959), consistent with its distinctive gait patterns.

%%%%%%%%%%%%%%%%%%%%%%%%%%%%%%%%%%
% Discussion and Conclusion
%%%%%%%%%%%%%%%%%%%%%%%%%%%%%%%%%%
\FloatBarrier

\section{Discussion and Conclusion}
\label{sec:conclusion}

We proposed a multimodal fusion framework integrating skeleton dynamics and clinical gait parameters for video-based CP severity assessment. Key findings include: (1)~fusion consistently outperforms the skeleton-only baseline, with the largest gain at Level~III (+11.9\% recall) where gait deviations are most salient; (2)~Grad-CAM-guided feature curation outperforms exhaustive feature sets, confirming that domain-informed selection reduces noise~\cite{jun2023hybrid}; and (3)~simple concatenation outperforms cross-attention (+1.78\%), suggesting that on moderately sized clinical datasets, additional parameterization risks overfitting.

By connecting model attention with explicit clinical gait descriptors, our framework enables interpretable predictions that clinicians can verify against established biomechanical knowledge~\cite{samadi2024machine,xiang2025explainable}. Limitations include severe under-representation of GMFCS Level~IV ($N_{\text{test}}{=}14$), reliance on 2D pose coordinates (3D lifting~\cite{jing2023deep} could help), and evaluation on a single backbone (extending to CTR-GCN~\cite{chen2021channel} would strengthen conclusions). In summary, combining skeleton dynamics with Grad-CAM-guided clinical gait features achieves 70.86\% accuracy (+5.6~pp over baseline) with clinically meaningful feature importance.

%%%%%%%%%%%%%%%%%%%%%%%%%%%%%%%%%%
% Acknowledgments
%%%%%%%%%%%%%%%%%%%%%%%%%%%%%%%%%%
\section{Acknowledgments}
We thank the authors of the CP gait dataset~\cite{kidzinski2020deep}.

%%%%%%%%%%%%%%%%%%%%%%%%%%%%%%%%%%
% Bibliography
%%%%%%%%%%%%%%%%%%%%%%%%%%%%%%%%%%
%%%%%%%%%%%%%%%%%%%%%%%%%%%%%%%%%%
% Bibliography
%%%%%%%%%%%%%%%%%%%%%%%%%%%%%%%%%%
%%%%%%%%%%%%%%%%%%%%%%%%%%%%%%%%%%
% Bibliography
%%%%%%%%%%%%%%%%%%%%%%%%%%%%%%%%%%

%%%%%%%%%%%%%%%%%%%%%%%%%%%%%%%%%%
% Bibliography
%%%%%%%%%%%%%%%%%%%%%%%%%%%%%%%%%%

\end{document}